\documentclass[prc,reprint,nofootinbib]{revtex4-1}
\usepackage{graphicx}   
\usepackage{latexsym}   

\begin{document}
\title{Sub-threshold charm production in nuclear collisions}

\author{J. Steinheimer$^1$, A. Botvina$^{1,2,3}$ and M. Bleicher$^{1,2,4}$}

\affiliation{$^1$ Frankfurt Institute for Advanced Studies, Ruth-Moufang-Str. 1, 60438 Frankfurt am Main, Germany}
\affiliation{$^2$ Institut f\"ur Theoretische Physik, Goethe Universit\"at Frankfurt, Max-von-Laue-Strasse 1, D-60438 Frankfurt am Main, Germany}
\affiliation{$^3$Institute for Nuclear Research, Russian Academy of Sciences, 117312 Moscow, Russia}
\affiliation{$^4$John von Neumann Institute for Computing (NIC), Forschungszentrum J\"ulich, 52425 J\"ulich, Germany}

\begin{abstract}
We present the first predictions for sub-threshold open charm and charmonium production in nuclear collisions. The production mechanism 
is driven by multi-step scatterings of nucleons and their resonance states, accumulating sufficient energy for the 
production of $J/\Psi$ and $\Lambda_c + \overline{D}$. Our results are of particular importance for the CBM experiment at FAIR,
as they indicate that already at the SIS100 accelerator one can expect a significant number of charmed hadrons to be produced.
This opens new possibilities to explore charm dynamics and the formation of charmed nuclei.
\end{abstract}


\maketitle

Charmed hadron production is considered to be an excellent probe of the properties
of hot and dense nuclear matter. Early works have argued that charmonium suppression
in central nuclear collisions may serve as signal for the formation of a deconfined medium,
the so called quark gluon plasma (QGP) \cite{Matsui:1986dk}. As the charm quark mass is much higher than the typical scale of QCD, 
charm in the traditional scenario is only produced in the very early stages of a nuclear collision where relative momenta are still large. 
In essence, it serves as a messenger of the properties of that stage.
A focus of recent investigations was on charm production at ultra-relativistic energies,
i.e. in experiments at the LHC and RHIC accelerators (see e.g. \cite{vanHees:2007me,Rapp:2009my,He:2012df,Aichelin:2012ww,Uphoff:2011ad,Uphoff:2012gb,Young:2011ug,Vitev:2007jj,Gossiaux:2010yx,Gossiaux:2012th,Lang:2012cx}).\\
It is expected to be even more interesting to study charm production at lower energies, for
example at the planned FAIR facility. At such low energies the system created is close to 
the transition between the hadronic phase and the QGP at very high net baryon densities.
Therefore the charm quarks and hadrons will be born in a very strongly interacting system of high baryon density,
opening up the possibility to study charm interactions with cold and hot hadronic matter.
In the physics program of the CBM experiment at FAIR the study of open charm and charmonium 
plays an essential role \cite{Friman:2011zz}. However, the FAIR project is planned to start with the SIS100 accelerator,
which will be able to accelerate a beam of heavy ions only to an energy of $E_{\mathrm{lab}}=11$ A GeV, an energy
which is below the charm production threshold in elementary collisions. 
In order to verify if the planned CBM experiment at the FAIR facility
is fit to do studies on open charm and charmonium production, it is of
great importance to have reliable estimates on the production cross sections these states.
In this paper will provide such estimates for sub-threshold charm production in nuclear collisions,
meaning charm production at beam energies below their elementary p+p threshold.\\
In the first part of the paper we will introduce the model which we will use and the mechanism employed
for charm production. In the second part we will show our results on open charm, charmonium and charmed nuclei
production at the SIS100 accelerator and in the final part we will discuss our results and their relevance for
the planned experiments at FAIR.

\begin{table}[b]
\centering
\begin{tabular}{|c|c|}
  \hline
Process & Threshold Energy [GeV] \\ \hline
$N+N \rightarrow N+N+J/\psi$ & $4.973$ \\ \hline
$N+N \rightarrow N+\Lambda_c+\overline{D}$ & $5.096$ \\ \hline
$N+N \rightarrow N+N+D+\overline{D}$ & $5.611$ \\ 
  \hline
		\end{tabular}
		\caption{Threshold center-of-mass energies, as implemented in UrQMD, for different charm production processes. $N$ refers to any ground state nucleon.}\label{t1}
\end{table}

\begin{table}[t]
\centering
\begin{tabular}{|c|c|c|c|}
  \hline
Name & Mass [GeV] & Width [GeV] & Spin \\ \hline
N*(2600) & 2.600  & 0.65 & $11/2$ \\ \hline
N*(2700) & 2.700  & 0.40 & $13/2$ \\ \hline
N*(3100) & 3.100  & 1.30 & $15/2$ \\ \hline
N*(3500) & 3.500  & 1.30 & $17/2$ \\ \hline
N*(3800) & 3.800  & 1.30 & $17/2$ \\ \hline
N*(4600) & 4.600  & 1.30 & $19/2$ \\ \hline
$\Delta$*(2420) & 2.420 & 0.40 & $11/2$ \\ \hline
$\Delta$*(2750) & 2.750 & 0.40 & $13/2$ \\ \hline
$\Delta$*(2950) & 2.950 & 0.50 & $15/2$ \\ \hline
$\Delta$*(3300) & 3.300 & 1.00 & $17/2$ \\ \hline
$\Delta$*(3500) & 3.500 & 1.00 & $19/2$ \\ \hline
$\Delta$*(3700) & 3.700 & 1.00 & $19/2$ \\ \hline
$\Delta$*(4200) & 4.600 & 1.00 & $21/2$ \\
  \hline
		\end{tabular}
		\caption{Newly introduced baryonic resonances}\label{t2}
\end{table}

\section{Charm production}
Estimates and predictions on the charm production cross section in p+p and A+A collisions
are usually based on a perturbative approach of QCD (see e.g. \cite{Vogt:2001nh}). As one approaches the threshold energy required to produce charm in p+p collisions, as given in table \ref{t1}, one expects these perturbative methods to break down, especially in A+A collisions.
Previous work on near threshold charm production, in A+A collisions, is based on a parametrized p+N cross section \cite{Cassing:2000vx,Linnyk:2008hp} or an effective interaction Lagrangian \cite{Liu:2003be,Rekalo:2002wg}. In such a study all charm is essentially produced in the first binary collisions, leading to binary scaling of charm production, thus the main contribution to sub-threshold production is from the Fermi momenta of the nucleons, shifting the threshold.
However it was already shown that deep below the threshold, the role of Fermi momenta is negligible \cite{Steinheimer:2016vzu} and one should expect a breakdown of the strict scaling of charm production with the number of binary collisions because the energy per individual p+p collision is not sufficient to produce a charm quark pair, i.e. that secondary baryon interactions are essential for particle production at and below the elementary production threshold. Therefore one
expects contributions of non-perturbative processes to the charmed hadron cross section.
One such a process could be the creation of charm in the decay of a heavy excited state, e.g. a baryonic resonance. Such a process has already shown to successfully describe near threshold production of the $\phi$ mesons \cite{Steinheimer:2016vzu,Steinheimer:2015sha}. Here the main contribution to the large sub-threshold $\phi$ multiplicity are the secondary scatterings of produced heavy baryonic states with other incoming baryons. It was found that these states decay and populate the multiplicity distributions of produced hadronic species according to phase space dominance. One should note that this is in essence nothing else but the application of Fermi's ''golden rule'' which states that in a decay process the final state's population probability is proportional to a (constant) matrix element (which has to be fixed) and the available phase space density for the process. It is known that such decays are the dominant process in associated $\Lambda$+K production (see e.g. \cite{Agakishiev:2014dha}). Here, we include and explore these processes for the first time, to make quantitative predictions on the sub-threshold production of charmed hadrons in nuclear collisions.

We will employ the UrQMD transport model \cite{Bass:1998ca,Bleicher:1999xi}, which already includes a extensive list of baryonic resonances. In order to describe charm production we extend the list of baryons to higher masses. As baryonic resonances act as an intermediate energy reservoir we will additionally include all known nucleon and Delta resonance with masses up to 4.6 GeV \cite{Agashe:2014kda}, since only these high mass states may decay into a $J/\Psi$ or $\Lambda_c + \overline{D}$ or $N + D + \overline{D}$.

The direct resonance production cross section, in elementary p+p collisions, implemented in UrQMD
follows from a phenomenological parametrization of measured experimental cross sections and phase space considerations. 
Here the cross section has the general form:
\begin{equation}\label{crosspp}
	\sigma_{1,2 \rightarrow 3,4}(\sqrt{s}) \propto (2 S_3 +1)(2S_4 +1) \frac{\left\langle p_{3,4} \right\rangle}{\left\langle p_{1,2} \right\rangle} \frac{1}{s} \left| M(m_3,m_4) \right|^2
\end{equation}
where $S_3$ and $S_4$ are the spin of the outgoing particles, and $\left\langle p_{i,j} \right\rangle$ the average momentum of the in- and out-going particles respectively. The matrix element $ \left| M(m_3,m_4) \right|$ is assumed to not explicitly depend on the spins but only on the masses of the outgoing particles. It is given as:
\begin{equation}
	\left| M(m_3,m_4) \right|^2 = A \frac{1}{(m_4-m_3)^2 (m_4+m_3)^2}
\end{equation}

\begin{figure}[t]	
\includegraphics[width=0.5\textwidth]{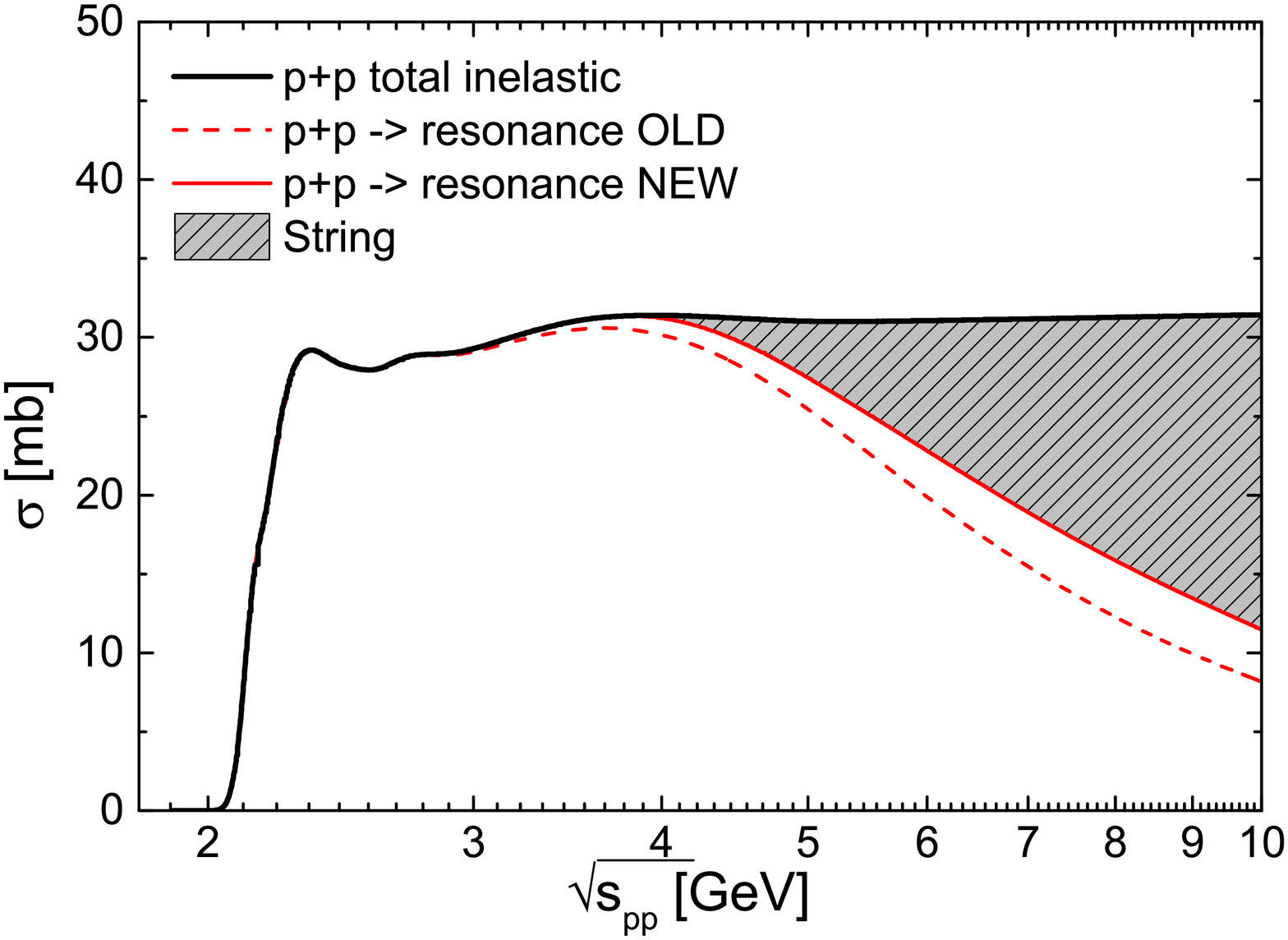}	
\caption{[Color online] Total inelastic cross section of p+p collisions, as implemented in UrQMD (black solid line). The red line depicts the part of the p+p cross section which corresponds to the excitation of at least one baryonic resonance.}\label{f1}
\end{figure}		

The parameter $A$ is determined by a fit to available data and is the same for any specific process, i.e. $N+N \rightarrow N+N$*, for all $N$*. The total inelastic p+p cross section from resonance excitation then follows as the sum of all possible channels (\ref{crosspp}). The mass dependence of the production cross section is then essentially determined by phase space, and not the parameter A, which has been shown to lead to a good description of measured resonance production cross sections \cite{Bass:1998ca}.
The resulting inelastic cross sections of p+p collisions, in the model, are shown in figure \ref{f1}. Here we compare the old resonance contribution (red dashed line) with the new implementation which includes the above mentioned heavy states (red solid line). At low energies this cross section essentially makes up for the total inelastic p+p cross section implemented in the model. As the resonance contribution to the total inelastic cross section increases, the contribution of the string must decrease at the same time in order to conserve the total inelastic cross section. All the before mentioned assumptions can also be used to directly compute any $N+N$ resonance contribution to the inelastic cross section, where $N$ can be either a proton or neutron. We can therefore directly apply our model to collisions of heavy nuclei containing protons as well as neutrons. 
The heaviest baryonic states, which eventually decay into the open and hidden charm states, will not be excited in the initial binary scatterings which are governed by the cross section shown in figure \ref{f1} but originate from secondary interactions of initially produced baryonic resonances which serve as an intermediate energy reservoir.   

\begin{figure}[t]	
\includegraphics[width=0.5\textwidth]{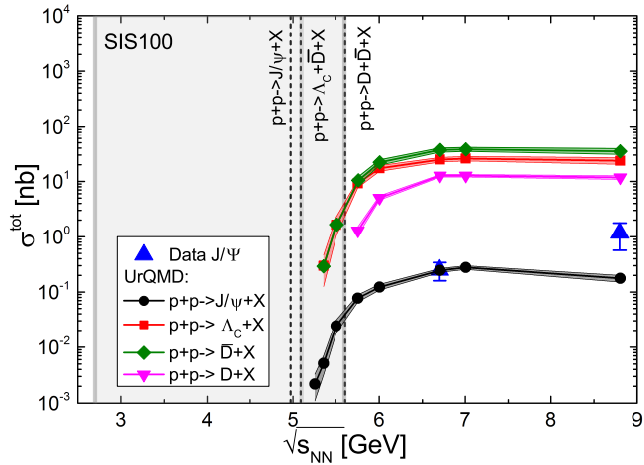}	
\caption{[Color online] $J/\Psi$, $\Lambda_c$, $\overline{D}$ and $D$ production cross section, from UrQMD, in p+p collisions as function of the collision energy. Experimental measurements of the $J/\Psi$ cross section are indicated as symbols \cite{Bamberger:1977mr,Corden:1980ht,Abt:2005qr}. The threshold energies of the corresponding channels are also indicated as vertical lines. The grey area corresponds to the beam energy range expected for heavy ion collisions at the SIS100 accelerator.}\label{f2}
\end{figure}

After fixing the production probability of the heavy resonances one needs to determine the branching fraction of the N* into the relevant charm channels, i.e. we need to determine the probability of N*$\rightarrow \mathrm{N} + J/\Psi$, N*$\rightarrow \Lambda_{c} + \overline{D}$ and N*$\rightarrow  \mathrm{N} + D + \overline{D}$. If $\Gamma_{tot}$ is the total decay width of a given resonance and $\Gamma_{FS}$ is the partial decay width
of that same resonance decaying into a particular final state FS, then the branching fraction $\Gamma_{\mathrm{FS}}/\Gamma_{tot}$ is defined as the fraction of resonances which decay to a certain final state FS. These partial decay widths are the essential free parameters of our approach and are not calculated explicitly.
To fix this crucial input we use experimental data on the measured $J/\Psi$ cross section in p+p collisions at $\sqrt{s_{pp}}=6.7$ GeV, which is shown in figure \ref{f2}. Note that the
p+p cross section in \cite{Bamberger:1977mr} is actually deduced from $J/\Psi$ production from protons on a Hydrogen target. Furthermore we assume that the $J/\Psi$ can only be created in 
a N* decay (it is forbidden in the $\Delta$* decay). The resulting branching fraction, to describe the experimentally measured $J/\Psi$ yield is $\Gamma_{\mathrm{N}+J/\Psi}/\Gamma_{tot} = 5 \cdot 10^{-5}$. This branching ratio is two orders of magnitude smaller than the corresponding decay into a $\phi$ meson. The resulting, energy dependent $J/\Psi$ cross section is shown in figure \ref{f2} as black line. We have also checked that the model reproduces the original measurement of the cross section of proton induced $J/\Psi$ production on Beryllium at a beam energy of $E_{\mathrm{lab}}= 23$ A GeV, which is of the order of 1-2 nanobarn \cite{Aubert:1974js} (with very large statistical and systematic errors).

The threshold for $\Lambda_c + \overline{D}$ creation is only slightly larger than that of the $J/\Psi$. It has been shown that the relative importance of associated charm production near the threshold is significant \cite{Andronic:2007zu}. 
In our case the relative branching ratio, of N*$\rightarrow \Lambda_c+ \overline{D}$ to N*$\rightarrow$N+$J/\Psi$, is taken from the statistical approach to charmed hadron production \cite{Andronic:2007zu}. In this approach all possible decay channels which may lead to a $\Lambda_c$ have been taken into account, for example the $\Sigma_c \rightarrow \Lambda_c +\pi$. In a sense we therefore take into account all possible contributions to the $\Lambda_c$ in this one effective parameter, the $\Lambda_c$ to $J/\Psi$ relative production rate.
In the statistical model, at the lowest beam energy measured $\sqrt{s_{pp}}=6.7$ GeV, about half of the expected charm will be in form of a $\Lambda_c$ while the $J/\Psi$ yields is about 100 times smaller. Therefore we have introduced the decay:
N*$\rightarrow \Lambda_c + \overline{D} $ with a branching fraction of $\Gamma_{\Lambda_C+\overline{D}}/\Gamma_{tot} = 1 \cdot 10^{-2}$.
Note that at the moment we have not included explicitly a possible decay of the $\Delta$* to $\Sigma_c \rightarrow \Lambda_c +\pi$. 
Finally we have also included the N*$\rightarrow  \mathrm{N} + D + \overline{D}$ branching fraction from statistical relative abundances as $\Gamma_{\mathrm{N} + D +\overline{D}}/\Gamma_{tot} = 2 \cdot 10^{-3}$.
The resulting production cross sections are shown in figure \ref{f2}. Note that the associated production cross-section is smaller than early estimates from a hadronic Lagrangian model \cite{Liu:2003be}. In the mentioned model the cross section is actually calculated from an effective interaction Lagrangian and not deduced from phase space considerations as in our approach. However, since our cross section is smaller we believe that our results can be considered a conservative estimate.

\begin{figure}[t]	
\includegraphics[width=0.5\textwidth]{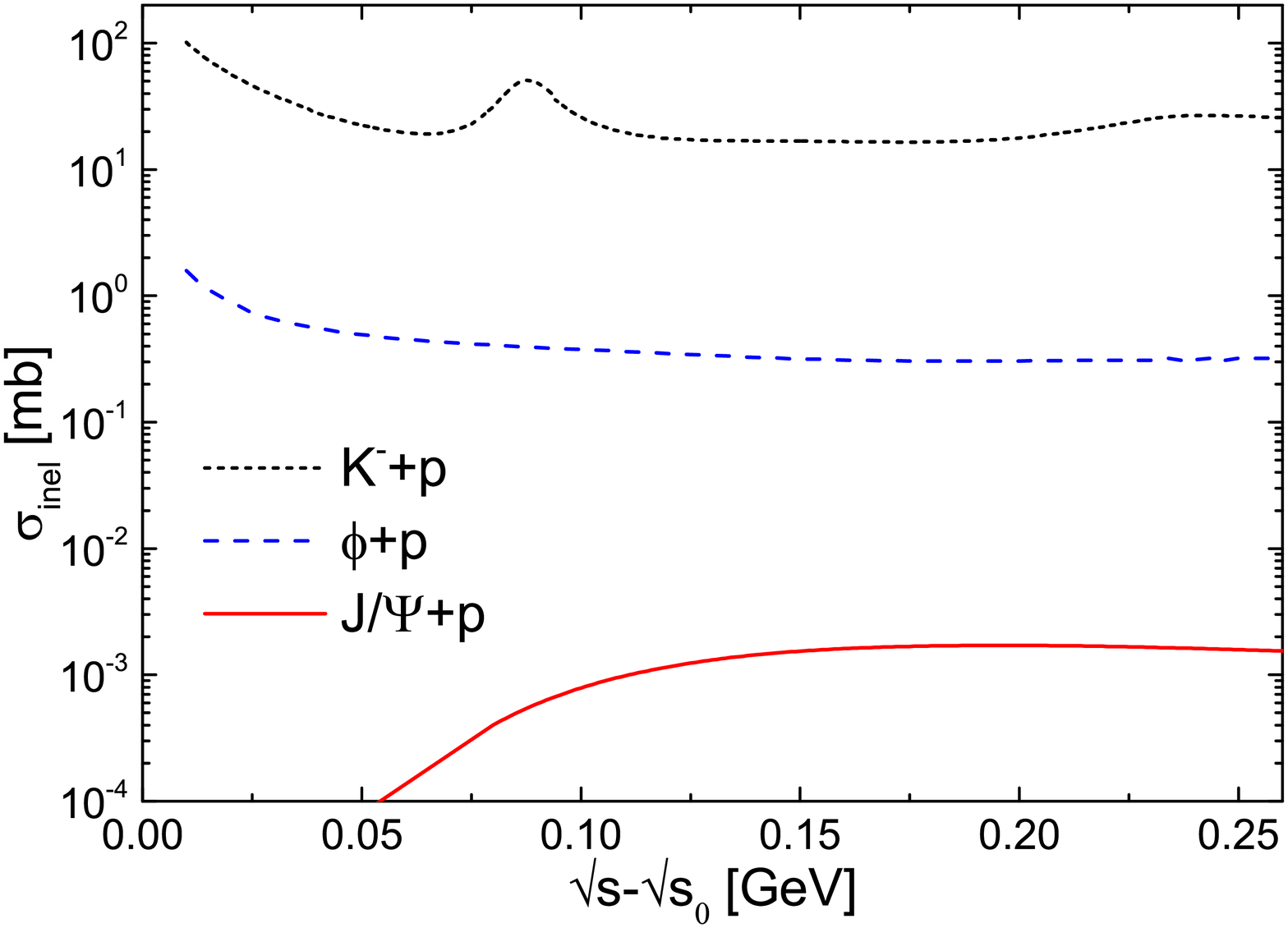}	
\caption{[Color online] Total inelastic $J/\Psi + p$ cross section as function of center of mass energy from detailed balance relations. The $J/\Psi$ cross section is compared to inelastic $\phi$ and $K^-$ scattering of the proton also from detailed balance relations. The rich structure in the $K^-$ + p scattering cross section is due to number of hyperonic resonances which can be excited in that channel (the peak corresponds mainly to the excitation of the $\Lambda$(1520)). }\label{f4}
\end{figure}		

\begin{figure}[t]	
\includegraphics[width=0.5\textwidth]{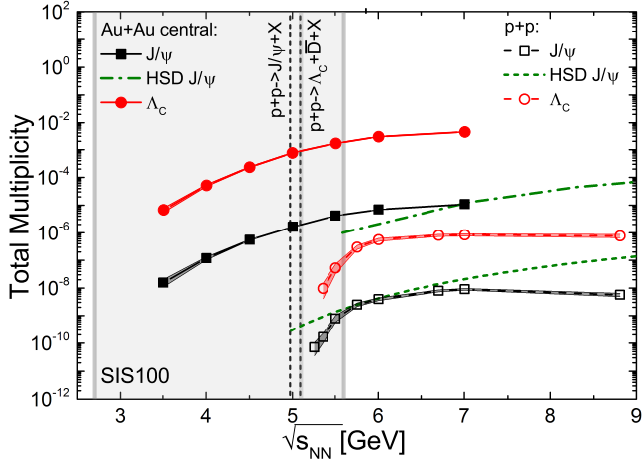}	
\caption{[Color online] Production yields of $J/\Psi$ and $\Lambda_C$ in p+p and central Au+Au reactions as a function of the collision energy. We compare our results with previous HSD model predictions \cite{Linnyk:2008hp}. The threshold energies of the corresponding channels in p+p reactions are again indicated as vertical lines. The grey area corresponds to the beam energy range expected for heavy ion collisions at the SIS100 accelerator.}\label{f3}
\end{figure}		

One can clearly see that the choice of our branching fraction fits the data point at $\sqrt{s_{pp}}=6.7$ GeV and nicely shows the correct threshold behavior. Due to the drop of the resonance excitation cross section at higher beam energies (the lack of the string cross section in our study), also the $J/\Psi$ cross section eventually saturates and will drop at even larger energies, which is why we cannot describe charm production at higher energies. However, since we are only interested in sub-threshold production we can safely ignore the contribution to charm production from even higher energy strings which would contribute for example to better describe $J/\Psi$ production at the higher beam energy of $\sqrt{s_{\mathrm{NN}}}=8.7$ GeV, shown in figure \ref{f2} \cite{Corden:1980ht}.
Consequently, we also did not include pair production of $D+\overline{D}$ from the string excitation, as the corresponding string threshold energy is larger than that of the resonance processes. Since we are only interested in the sub-threshold production process where string excitations are of minor importance.
Furthermore we do not take into account rescattering and absorption processes of the charmed hadrons in the hadronic medium. At the moment we are only interested in the total production cross section which is not influenced much by hadronic rescatterings (charmed hadrons have a very small inelastic rescattering cross section).

Including the production of the $J/\Psi$ via the excitation and decay of a heavy baryonic state has an additional advantage. We can use detailed balance relations to extract, from the earlier determined branching ratio, the total inelastic cross section of $J/\Psi + p \rightarrow N^* \rightarrow X $. The resulting total inelastic cross section as a function of free invariant mass $\sqrt{s}-\sqrt{s_0}$, with $\sqrt{s_0}=m_{J/\Psi}+m_N$ in case of the $J/\Psi$, is shown in figure \ref{f4}. Here it is also compared to the total inelastic cross sections of $K^- + p$ and $\phi +p$. We find that the $J/\Psi$ has a very small cross section with the nucleon, on the order of a few micro barn, as expected for near threshold production from short range QCD calculations \cite{Kharzeev:1994pz}. This result does also support the omission of the inelastic hadronic rescattering for the $J/\Psi$.

\section{Results}

\begin{figure}[t]	
\includegraphics[width=0.5\textwidth]{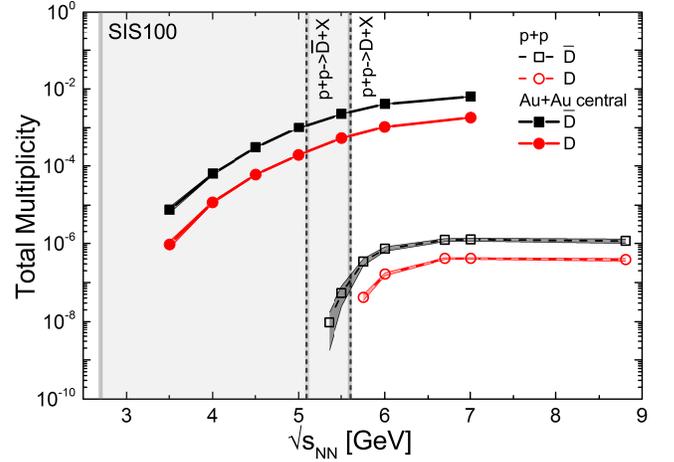}	
\caption{[Color online] Production yields of $D$ and $\overline{D}$ mesons in p+p and central Au+Au reactions as a function of the collision energy. The threshold energies of the corresponding channels in p+p reactions are again indicated as vertical lines. The grey area corresponds to the beam energy range expected for heavy ion collisions at the SIS100 accelerator.}\label{f6}
\end{figure}		

Figures \ref{f3} and \ref{f6} summarize the results of the sub-threshold charm production in nuclear collisions, obtained with the above described model. As expected the production multiplicity for charmed hadrons does not show the strong threshold behavior, as observed in the p+p reactions, in the Au+Au collisions. In the model, due to secondary interactions,
there essentially is no sharp production threshold anymore and the $J/\Psi$ can be produced over a wide range of beam
energies even below the elementary threshold. Again we want to make clear that the heavy resonance which eventually decays into the charm hadrons cannot be produced
in a first binary p+p collision (deep) below the elementary threshold, even in A+A. It may however be produced in a secondary collision
of e.g. two already excited states which lead to a sufficiently high $\sqrt{s}$ to produce the high mass resonance.
We would like to point out that the threshold for the $D+\overline{D}$ production is also smeared, due to secondary interactions, in 
A+A collisions. However we observe that the associated production $\Lambda_c + \overline{D}$ dominates the
$\overline{D}$ yield which can be observed in figures \ref{f2}, \ref{f3} and \ref{f6}.
At a fixed target beam energy of $E_{\mathrm{lab}}=6$ A GeV we expect a yield of $1 \times 10^{-7}$ $J/\Psi$, $5 \times 10^{-5}$ $\Lambda_c$, $6 \times 10^{-5}$ $\overline{D}$ and $1 \times 10^{-5}$ $D$ per central Au+Au event. In order to estimate the expected numbers of measured charmed hadrons from experiment, a dedicated study on reconstruction efficiencies and experimental acceptance is in order. This highlights the need for reliable predictions, as supplied from this paper, on expected charmed hadron yields.
For the highest available beam energy for heavy nuclei at the SIS100, $E_{\mathrm{lab}}=11$ A GeV, we expect that yield to be one order of magnitude larger. Hence we predict that a significant number of charmed hadrons can be measured at the CBM experiment, already with the SIS100 accelerator in place. This is of particular interest as the baryon number densities achieved at these low beam energies are very large, opening up the possibility to study charm production and propagation in a very dense hadronic environment. 
Of particular interest here is to understand and quantify the interaction of charmed hadrons with the nuclear environment and possible effects of chiral symmetry restoration of charmed hadron properties.

A specific example of how the interactions of charmed hadrons can be inferred would be the measurement
of charmed nuclei. The existence of bound states of nucleons and one or more $\Lambda_c$ was first
suggested in \cite{Tyapkin,Dover:1977jw}. Although their actual binding energy are still under discussion
it is generally agreed that such states should exist \cite{Tsushima:2002ua,Miyamoto:2016hqo}. Consequently the
detection and therefore confirmation of the existence of bound nuclear states with charm would be an
extremely important discovery. 
We propose that the CBM experiment is best suited for the production and detection of
charmed nuclei, due to the high baryon density created and the expect high statistics 
measurements.
Within the above discussed UrQMD model we can estimate the expected number of charmed nuclei produced
in minimum bias collisions.
It has been shown that in central nuclear collisions  
the coalescence mechanism, which assembles light nuclear clusters from the 
produced baryons is important and it can be used to predict the yields of
 fragments like nuclei and hypernuclei with reasonable accuracy \cite{Ste12,star,alice}. 
Previously we have developed the Coalescence of Baryons (CB) model 
\cite{Bot15} which was successfully used to calculate the yields of hypernuclei,
i.e. nuclei with strangeness. In this work, we have extended this approach, to include the 
coalescence of charmed $\Lambda_c$ particles and nucleons to form charmed 
light clusters. 
The coalescence model, as described in \cite{Bot15}, is realized in both the 
coordinate and momentum (velocity) space. It is assumed that 
baryons contained in a cluster with mass number $A$ can only be located 
within a radius of 2.0$A^{1/3}$fm from the center of the cluster. This is 
consistent with the nuclear density extracted from secondary decays 
of excited nuclei. The parameter $v_C$ represents effectively the 
maximum difference in relative velocities between nucleons within 
the same coalescent cluster.
The uncertainty in the 
binding energy of such nuclei can be evaluated by the variation of the 
coalescent parameters $v_C$, similar as it is applied for normal clusters and 
hyper-clusters. The largest $v_C$=0.22 corresponds approximately 
to the Fermi motion in large nuclei. The value around $v_C$=0.10 was 
obtained previously from the coalescent description of 
experimental data on normal light cluster production ($A \le 4$) in 
Heavy Ion collisions at energies around $1-10$ A GeV (see \cite{Ste12}).
Also, this corresponds to internal excitation energies (calculated from the relative motion 
of nucleons) of light clusters around several MeV, consistent 
with their binding energy.

\begin{figure}[t]	
\includegraphics[width=0.5\textwidth]{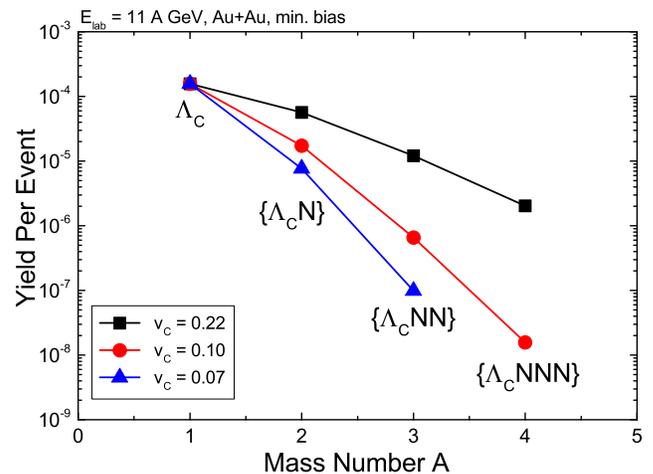}	
\caption{[Color online] Production yields of light, single charmed nuclei as function of the mass number, for different values of the coalescence parameter $v_C$. We show results for min. bias collisions of Au+Au nuclei at a fixed target beam energy of $E_{\mathrm{lab}}= 11$ A GeV.}\label{f5}
\end{figure}		

In Fig.~\ref{f5} we show the yields (per minimum bias Au+Au collision at a fixed target beam energy of $E_{\mathrm{lab}}= 11$ A GeV) of $\Lambda_c$ and 
light clusters, of different mass number $A$, containing $\Lambda_c$ obtained within our approach. 
We have considered 3 different coalescence parameters for the coalescence 
stage, which reflect different assumptions on the binding energy of the $\Lambda_c$. The largest 
$v_c$=0.22 corresponds to large relative velocities between the baryons in 
clusters. For small clusters 
this can be considered a very optimistic estimate. The $v_c$=0.1 is the most reasonable choice, 
since this parameter choice leads to a good description of the yields of normal light nuclei in nuclear collisions \cite{Ste12}. 
We also show the results for an even smaller parameter 
$v_c$=0.07 in order to give an conservative estimate on the lower limit for charmed 
nuclei production. 

As charmed-nuclei can be identified through their mesonic decays (see e.g. \cite{Ghosh:2016jjv,Yao:2006px}),
we think is it feasible to investigate their decay
into normal light clusters and hyperons. 
For example, a system with A=3 may decay via the production of 
an intermediate hyper-nucleus :
$\Lambda_c np \rightarrow \Lambda np + \pi^{+} \rightarrow ^{3}He + 
\pi^{-} + \pi^{+}$. The decay of $\Lambda np$ nucleus was already 
reliably observed in relativistic heavy ion collisions \cite{star}. 
Since sophisticated correlation measurements and the reconstruction of the 
invariant masses are necessary for their identification, it is most important to be able to provide the experiment
with model simulations, such as presented here, which can be used for further feasibility studies for the CBM detector.

\section{Summary}

We provide pioneering estimates on the deep sub-threshold production of charmed hadrons in nuclear collisions. We find that the CBM experiment is well suited to make successful measurements on $J/\Psi$, $\Lambda_c$, $D$and $\overline{D}$ production yields and even spectra, at fixed target beam energies as low as $E_{\mathrm{lab}}= 6$ A GeV. This finding is very important as it enables studies of charmed hadron properties at very high baryon densities with the SIS100 accelerator. Furthermore we propose that the CBM experiment may be able to, for the first time, confirm the existence of bound states of charmed baryons with nucleons, so called charmed-nuclei. Such a finding would constitute a remarkable step in understanding the interactions of charmed hadrons with normal nuclear matter.

\section{Acknowledgments}
The authors thank Michael Deveaux for helpful discussions about the CBM experiment. 
This work was supported by GSI and BMBF. The computational resources were provided by the LOEWE Frankfurt Center for Scientific Computing (LOEWE-CSC).


\end{document}